\documentclass[prd,twocolumn,showpacs]{revtex4}
\def\b{\bar}
\def\d{\partial}

\def\cD{{\cal D}}

\def\m{\mu}
\def\n{\nu}

\def\t{\tau}

\def\~{\widetilde}

\def\bY3{\bar Y_{,3}}
\def\Y3{Y_{,3}}
\def\z{\zeta}
\def\Z{{\b\zeta}}
\def\Y{{\bar Y}}
\def\cZ{{\bar Z}}
\def\`{\dot}
\def\be{\begin{equation}}
\def\ee{\end{equation}}
\def\bea{\begin{eqnarray}}
\def\eea{\end{eqnarray}}

\def\fn{\footnote}

\def\cF{{\cal F}}

\def\olam{\stackrel{\circ}{\lambda}}
\def\oX{\stackrel{\circ}{X}}

\def\mn{{\mu\nu}}

\usepackage{amsmath}
\usepackage{epsfig}

\begin{document}

\title{The Dirac -- Kerr-Newman electron.}

\author{Alexander Burinskii}
\affiliation{Gravity Research Group, NSI, Russian Academy of
Sciences,
 B. Tulskaya 52, 115191 Moscow, Russia}

\begin{abstract}
  We discuss the relation of the Kerr-Newman spinning particle to
the Dirac electron and show that the Dirac equation may naturally
be incorporated into Kerr-Schild formalism as a master equation
controlling the Kerr-Newman geometry. As a result, the Dirac
electron acquires an extended space-time structure of the
Kerr-Newman geometry  - singular ring of the Compton size and
twistorial polarization of the gravitational and electromagnetic
fields.

Behavior of this Dirac -- Kerr-Newman system in the weak and
slowly changed electromagnetic fields is determined by the wave
function of the Dirac equation, and is indistinguishable from the
behavior of the Dirac electron. The wave function of the Dirac
equation plays in this model the role of an ``order parameter''
which controls dynamics, spin-polarization and twistorial
structure of space-time.
\end{abstract}

\maketitle

\section{Introduction}

The discovery of the Kerr and Kerr-Newman solutions
\cite{Ker0,KNew} had  fundamental importance for all the areas of
theoretical physics from cosmology and astrophysics to superstring
theory. Interest to these solutions  in the physics of elementary
particles has been raised recently by the conjecture that black
holes may be created in the laboratory conditions.

On the other hand, the treatment of the Kerr-Newman solution as a
model of electron have been considered many times
\cite{Car,Isr,DKS,Bur0,IvBur,Bur1,IvBur1,Lop,Man,BurStr,BurOri,BurTwi,BurPra,ArkPer}
after the well known Carter remark \cite{Car} that the Kerr-Newman
solution has gyromagnetic ratio $g=2$ as that of  the Dirac
electron. If this coincidence is not occasional, there appears a
fundamental question -  what is the relation of the Dirac equation
to the structure of Kerr-Newman solution?

The Dirac equation is also the discovery of fundamental
importance. Practically, most of the theoretical and experimental
results in the particle physics are based on the Dirac equation
and its consequences. Nevertheless, the Dirac description of
electron can not be considered as complete for two reasons:

-it does not take into account gravitational field of
 electron,

- the Dirac wave function caries a very obscure information on the
space-time position and, especially, on the structure of electron.

Forming the wave packets one can only show that electron cannot be
localized inside the Compton region. Meanwhile, the multi-particle
QED description showed that the ``naked'' electron is point-like
 and the electron ``dressed'' by virtual photons  is smeared over
 the Compton region. The point-like structure of electron is also
 confirmed by the deep inelastic scattering. There were a lot of
 attempts to describe the extended structure of spinning particle on
 the base of field models, and some of them were related to gravity.

The Kerr solution gives us a natural description of  spinning
particle with gravity, and moreover, it hints us about the
relation to electron by the double gyromagnetic ratio
\cite{Car,DKS}, by the extension on the Compton region
\cite{Isr,Bur0,Lop}, by the wave properties \cite{Bur0}, and by
the reach spinor and twistorial structures
\cite{IvBur,Bur1,BurStr,BurOri,BurTwi,BurPra}.

The aim of this paper is to consider a model which sets a relation
between spinor solutions of the Dirac equation and the spinor
(twistorial) structure of the Kerr-Newman solution.
 In this paper we suggest a new approach  which is based on
the assumption that {\it the Dirac equation is complimentary to
the Kerr-Newman spinning particle and plays the role of a master
equation controlling the motion and orientation of the Kerr-Newman
twistorial space-time structure.} In this model the Dirac spinors
are matched to the spinor (twistorial) structure of the
Kerr-Newman spinning particle, and the Dirac wave function plays
the role of an order parameter controlling this system.

On the other hand, since {\it the behavior of this combined Dirac
-- Kerr-Newman system turns out to be determined by the Dirac wave
function, it is undistinguishable from the behavior of the Dirac
electron,} which allows one to interpret this model in the frame
of one-particle quantum theory.

Plane of the work is following.  To realize our aim we have to
find a bridge between the Dirac theory and the Kerr-Newman
structure, i.e. to obtain some objects which are common for both
these structures.

In sec.II we consider the real and complex structures of the
Kerr-Newman geometry and obtain that the motion and polarization
of the Kerr-Newman spinning particle are controlled by two null
vectors $k_L$ and $k_R$ which are related to a ``point-like'' {\it
complex} representation of the Kerr-Newman geometry.

In sec.III, analyzing the structure of Dirac equation in the Weyl
basis, we obtain two similar null vectors $k_L$ and $k_R$ which
are determined by the Dirac wave function $\Psi$ and  control the
dynamics and polarization of the Dirac electron.

It allows as to set a link between the Dirac and Kerr-Newman
spinning particle, which is performed in sec.IV. As a result, the
the Dirac -- Kerr-Newman electron acquires a definite extended
space-time structure of the Kerr-Newman geometry, and the Dirac
wave function takes the role of an ``order parameter'' which
controls polarization and dynamics of this structure.  In
particular, the Kerr-Newman twistorial structure acquires a
dependence on the vector potential of the external electromagnetic
field via the Dirac equation.

In the following sections we discuss the properties and
consequences of the Dirac -- Kerr-Newman model in comparison with
the properties of electron in Dirac theory and QED.

In particular, in sec.V we show that renormalization of the
self-energy of the Dirac -- Kerr-Newman electron may be performed
by gravity for different distributions of the mechanical mass and
charge, and we describe the approach which allows one to get the
regular ``dressed'' solutions for different values of charge, mass
and spin.

In sec.VI, following to the known approach to quantum fields in
curved spaces \cite{BirDev,deWit},  we consider regularization of
the stress-energy tensor, which has  a specific realization in the
Kerr-Newman geometry.

In sec.VII, treating the obtained recently multi-particle
Kerr-Schild solutions \cite{Multiks}, we show that the Dirac --
Kerr-Newman electron model takes an intermediate position between
the one-particle Dirac electron and the multi-particle structure
of electron ``dressed'' by virtual photons in QED.

Sec.VIII contains a discussion on the space-time structure of
electron and the role of wave function in the Dirac theory, QED
and in the Dirac -- Kerr-Newman model.

In Conclusion we return to the origin of mass term in the Dirac
equation and discuss some possible ways for modification of this
model.

We use the spinor notations of the book \cite{WesBag} (see
Appendix A). However our spinors are commuting, which changes some
relations, and we give them in the Appendix B. Matching notations
to the paper \cite{DKS}, we use signature $(-+++)$,  Cartesian
coordinates $x^0=t, \quad x^\m=( t, x, y, z)$ and
$E=T_0^0=p_0=-p^0>0 $.

\section{Real and complex structures of the Kerr geometry}

{\bf The real structure of the Kerr-Newman geometry}, Kerr
congruence and the Kerr theorem were discussed many times, and we
refer readers to our previous papers, for example
\cite{BurOri,BurTwi,BurPra,Multiks,BurNst}.

Recall, that  angular momentum $J=\hbar /2$ for parameters of
electron is so high that the black hole horizons disappear and the
source of the Kerr-Newman spinning particle represents a naked
singular ring.

It was suggested in \cite{Bur0,IvBur} that the Kerr-Newman
singular ring represents a string which may have some excitations
generating the spin and mass of the extended particle-like object
- ``microgeon''. The Kerr-Newman singular ring is a focal line of
the principal null congruence which is a bundle of the lightlike
rays -- twistors. The null vector field $k^\m (x)$, which is
tangent to these rays, determines the form of metric \be g^\mn
=\eta^\mn + 2H k^\m k^n \label{KS} \ee (where $\eta^\mn$ is the
auxiliary Minkowski metric) and the form of vector potential \be
A_\m = {\cal A} (x) k_\m \label{Aem}\ee

for the charged Kerr-Newman solution.

The Kerr congruence is twisting (see figures in our previous works
\cite{BurOri,BurTwi,BurPra,BurNst}).  It is determined by the Kerr
theorem \cite{BurNst,Pen,KraSte} which constructs the Kerr-Schild
ansatz (\ref{KS}), skeleton of the Kerr-Newman geometry, as a
bundle of twistors \cite{BurTwi,BurPra,BurNst}.

{\bf Point-like complex representation of the Kerr-Newman
geometry. }

Complex representation of the Kerr-Newman geometry is important
for our treatment here since the complex Kerr-Newman source is
``point-like'', which corresponds to a ``naked'' electron in our
treatment.

Applying the complex shift $(x,y,z) \to (x,y,z+ia)$ to the source
$(x_0,y_0,z_0)=(0,0,0)$ of the Coulomb solution $q/r$, Appel in
1887(!) considered the resulting solution \be \phi(x,y,z)= \Re e \
q/\tilde r, \label{App}\ee
 where $\tilde r =\sqrt{x^2+y^2+(z-ia)^2}$ turns out to be
complex. On the real slice $(x,y,z)$, this solution acquires a
singular ring corresponding to $\tilde r=0.$ It has radius $a$ and
lies in the plane $z=0.$  The solution is conveniently described
in the oblate spheroidal coordinate system $r, \ \theta,$ where
$\tilde r =r+ia\cos\theta ,$ and one can see that the space is
twofolded having the ring-like singularity as the branch line.
Therefore, for the each real point $(t,x,y,z) \in {\bf M^4}$ we
have two points, one of them is lying on the positive sheet,
corresponding to $r>0$,  and another one lies on the negative
sheet, where $r<0$.

It was obtained that Appel potential corresponds  exactly to
electromagnetic field of the Kerr-Newman solution written in the
Kerr-Schild form \cite{Bur0}. The vector of complex shift $\vec
a=(a_x,a_y,a_z)$ corresponds to angular momentum of the
Kerr-Newman solution.

Newman and Lind \cite{New} suggested a description of the
Kerr-Newman geometry  in the form of a retarded-time construction,
where it is generated by a complex source, propagating along a
{\it complex world line} $\oX^\m(\t)$ in a complexified Minkowski
space-time $\mathbf{CM}^4$. The rigorous substantiation of this
representation can be achieved in the Kerr-Schild approach
\cite{DKS} which is based on the Kerr theorem and the Kerr-Schild
form of metric (\ref{KS}) which are related to the auxiliary
$\mathbf{CM}^4$ \cite{BurNst,BurKer,BurMag}. In the rest frame of
the considered Kerr-Newman particle, one can form two null
4-vectors $k_L=(1,0,0, 1)$ and $k_R=(1,0,0, -1),$
 and  represent the 3-vector of complex shift $i\vec a=i \Im m \oX^\m$ as the
  difference $i\vec a =\frac{ia}{2} \{ k_L -
k_R \}.$  The straight complex world line corresponding to a free
particle may be decomposed into the form

\be \oX^\m(\t) = \oX^\m(0) + \t u^\m + \frac{ia}{2} \{ k_L^\m -
k_R^\m \} ,\label{cwl}\ee

where the time-like 4-vector of velocity $u^\m=(1,0,0,0)$
 can also be represented via vectors $k_L$ and $k_R$

\be u^\m =\d _t \Re e \oX^\m(\t)=\frac 12 \{ k_L^\m + k_R^\m \}.
\label{umu} \ee
 It allows one to describe the
complex shift in the Lorentz covariant form.

One can form two complex world lines related to the complex
Kerr-Newman source, $\oX_{+}^\m(t+ia) = \Re e \oX^\m(\t) + ia
k^\m_L $ and $ \oX_{-}^\m(t-ia) = \Re e \oX^\m(\t) - i a k^\m_R .$
This representation will allow us to match the Kerr-Newman
geometry to the solutions of the Dirac equation.

Indeed, since the complex world line $\oX^\m (\t)$ is parametrized
by the complex time parameter $\t=t+i\sigma$, it represents a
stringy world sheet. However, this string is very specific, it is
extended along the imaginary time parameter $\sigma$.
 The world lines
$\oX_{+}^\m=\oX^\m(t+ia)$ and $\oX_{-}^\m=\oX^\m(t-ia)$ are the
end points of this open complex twistor-string
\cite{BurStr,BurOri,BurTwi,BurPra}, see fig.1. By analogue with
the real strings where the end points are attached to quarks, one
can conventionally call these complex point-like sources as
`quarks'.

\begin{figure}[ht]
\begin{center}
\centerline{\epsfig{figure=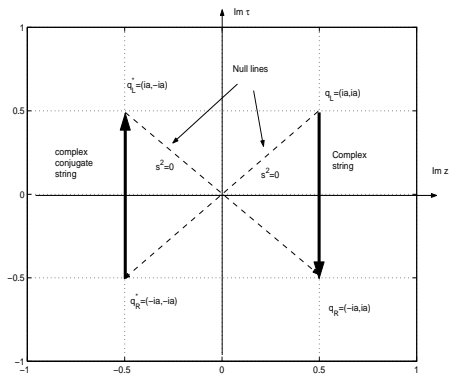,height=5cm,width=6cm}}
\end{center}
\caption{Positions of the complex point-like sources
(conventionally ``quarks'') at the ends of the open complex
Kerr-Newman string. The ``quarks'' form two pairs which are
related by the light-like intervals $s^2=0$. Two complex
conjugated strings are to be joined forming orientifold, i.e. a
folded closed string.}
\end{figure}
\section{ Dirac equation in the Weyl basis.}

In the Weyl basis the Dirac equation \be (\gamma^\m \hat \Pi _\m
+m)\Psi=0, \label{InitD} \ee where $\Psi = \left(\begin{array}{c}
 \phi _\alpha \\
\chi ^{\dot \alpha}
\end{array} \right),$
and $ \hat \Pi _\m = - i  \d _\m - e A_\m ,$

splits into the system \cite{BLP,Fei} \be
 \sigma ^\m _{\alpha \dot \alpha} (i \d_\m  +e A_\m)
 \chi ^{\dot \alpha}=  m \phi _\alpha , \quad
 \bar\sigma ^{\m \dot\alpha \alpha} (i \d_\m  +e A_\m)
 \phi _{\alpha} =  m \chi ^{\dot \alpha},
\label{Dir} \ee

The Dirac current \be J_\m = e (\bar \Psi \gamma _\m \Psi) = e
(\bar\chi  \sigma _\m  \chi + \bar\phi  \bar \sigma _\m  \phi ),
\ee where $\bar\Psi =(\chi ^+, \phi ^+ ),$ can be represented as a
sum of two lightlike components of opposite chirality \be J^\m_L =
e \bar\chi \sigma^\m \chi \ , \qquad J^\m_R = e \bar\phi
\bar\sigma^\m \phi. \ee

Forming  the null vectors $ k^\m_L = \bar\chi \sigma^\m \chi \ ,$
and $k^\m_R = \bar\phi \bar\sigma^\m \phi $, one can obtain for
their product\fn{We use the relations (\ref{s10}),(\ref{3m}) and
(\ref{s102}) for commuting spinors.}

\be k^\m_L k_{R \m}= (\bar\chi \sigma^\m \chi)(\bar\phi
\bar\sigma^\m \phi)= -2(\phi\bar\chi)(\chi\bar\phi)=
 2(\bar\chi\phi)(\bar\chi\phi)^+ \ .\ee

One can also form  two more vector combinations from the Dirac
spinors, $ m^\m = \phi \sigma^\m \chi \ ,$ and $\bar m^\m = (\phi
\sigma^\m \chi)^+= (\bar\chi \sigma^\m \bar\phi)$, which are
complex conjugated and have also the scalar product

\be  m^\m \bar m _\m = 2(\bar\chi\phi)(\bar\chi\phi)^+ . \ee

All the other products of  the vectors $k_L , \ k_R , m , \bar m $
are null.

The normalized four null vectors

\be n^a= \frac 1 {\sqrt{2(\bar\chi\phi)(\bar\chi\phi)^+}} (m ,
\bar m, k_L , k_R ), \quad a=1,2,3,4 , \ee  form {\it a field of
the natural quasi orthogonal null tetrad  which is determined by
the given solution $\Psi (x)$ of the Dirac equation.}

Notice, that the complex vectors $m$ and $\bar m$ turn out to be
modulated by the phase factor $\exp\{ 2 i p_\m x^\m \}$ coming
from the Dirac spinors and carry oscillations and also de Broglie
periodicity for the moving particle. Meanwhile, this phase factor
cancels for the real null vectors $k_L$ and $k_R$.

If $\Psi$ is a plane wave

\be \Psi = \left(\begin{array}{c}
 \phi _\alpha \\
\chi ^{\dot \alpha}
\end{array} \right) =
\left(\begin{array}{c}
 \breve{\phi} _\alpha \\
\breve{\chi} ^{\dot \alpha}
\end{array} \right) \exp \{i p_\m x^\m \},
\ee
 the Dirac equations take the form
 \be m \phi _\alpha  =
  -\Pi _\m \sigma ^\m _{\alpha \dot \alpha} \chi ^{\dot \alpha},\qquad
m \chi ^{\dot \alpha}  =
  -\Pi _\m \bar\sigma ^{\m \dot\alpha \alpha} \phi _{\alpha},
\label{DirSpl} \ee where \be \Pi_\m = p_\m -e A_\m .\label{Pi} \ee

Multiplying these equation by the spinors $\phi, \chi, \bar\phi,
\bar\chi$ to the left, one obtains the  null tetrad components of
$\Pi _\m$

 \be
\Pi _\m m^\m =- m \phi\phi = 0,\qquad \Pi _\m \bar m^\m = -  m
\chi\chi = 0. \label{mmcomp} \ee

 \be
\Pi _\m k_L ^\m =- m \bar\chi\phi ,\qquad \Pi _\m k_R ^\m = -
  m \bar\phi\chi.
\label{kkcomp} \ee

It shows that the vector $\Pi _\m$ is spanned by
 the real vectors $k_L$ and $k_R,$ i.e.
$ \Pi^\m = a k_L^\m + b k_R^\m \label{Pispan} \ ,$ where $
a=b=-\frac m {2(\phi\bar\chi)}. $ Therefore, \be \Pi^\m = -\frac
{m}{2\bar\phi\chi}(k^m_L +k^m_R ) \ . \ee In the rest frame \be
k^0_L =k^0_R , \quad k^i_L =-k^i_R \ ,\ee and  the space
components of $\Pi, \ (i=1,2,3)$ are cancelled.

The spin of electron is determined by the polarization vector
which has the form \cite{Fei} \be S^\m =i\overline\Psi \gamma^\m
\gamma^5 \Psi = k^\m_L - k^\m_R \ , \ee and therefore, in the rest
frame $S^0=0$ and $S=(0,S^i)$.

For normalized spinors  \be |\bar\phi \chi| = |\phi_1|^2 +
|\phi_2|^2=1 .\ee

For $E>0$ \be \bar\phi \chi = -1 , \ee which yields $\Pi^0 >0 .$
For $E<0$ \be \bar\phi\chi = 1 ,\ee
 and $\Pi^0 <0 .$

\section{Dirac equation as a master equation for the Kerr-Newman
twistorial structure}

In previous treatment we have seen that the complex Kerr-Newman
geometry is related to two null vectors $k_L$  and $k_R$ which
determine the momentum and angular momentum of Kerr-Newman
particle. We have also seen that the momentum and spin of the
Dirac electron in the Weyl basis are also expressed via two null
vectors $k_L$  and $k_R$. It allows us to connect the solutions of
the Dirac equation to twistorial structure of the Kerr-Newman
spinning particle by setting an equivalence for these null
vectors.

The Kerr-Schild ansatz for metric(\ref{KS}) is fixed by the null
vector field $k_\m (x) ,$ which is tangent to the Kerr principal
null congruence (PNC) and is determined by a complex function
$Y(x)$

\be k_\m dx^\m = P^{-1} (du + \bar Y d \zeta + Y d \bar\zeta - Y
\bar Y dv ), \label{Y} \ee

where $P(Y)$ is a normalizing factor and $\{u, v,\zeta,\bar \zeta
\} $ are the null Cartesian coordinates

\bea 2^{1\over2}\z &=& x + i y ,\qquad 2^{1\over2} \Z = x - i y ,
\nonumber\\ 2^{1\over2}u &=& z + t ,\qquad 2^{1\over2}v = z - t .
\label{ncc} \eea

The Kerr-Schild formalism is based on the null congruences which
are geodesic and shear-free.

{\bf The Kerr Theorem} \cite{Pen,KraSte} claims that all the
geodesic and shear-free congruences are determined by the function
$Y(x)$ which is a solution of the algebraic equation

\be F=0, \label{F} \ee

where the generating function $F$ is arbitrary holomorphic
function of the projective twistor variables

\be Y,\quad \lambda _{1} = \z - Y v,\quad \lambda _{2} = u + Y \Z
. \ee

Recall, that twistor is the pair $Z^a=\{ \psi_{\alpha}, \mu
^{\dot\alpha} \},$ where $\mu ^{\dot\alpha}= x^\m \bar\sigma _\m
\psi_\alpha ,$ and projective twistor is \be Z^a/\psi_1 =\{1 , Y,
\lambda _1 , \lambda _2 \}.\ee

Therefore, the target function $Y(x)$ is a projective spinor
coordinate $Y=\psi_2/\psi_1$, and function $F$ may be chosen as
homogenous function of $Z^a$.

The  complex world line $\oX _{+}^\m $ can
 be used as a complex Kerr-Newman source for  generating function $F$
 of the Kerr theorem.
 \be
 \oX _{+}^\m=\Re e \oX _{+}^\m + ia k^\m_L =\Re e \oX _{+}^\m
 +ia \bar\chi\sigma^\m \chi \label{csource}.
 \ee
Indeed, as it was shown in \cite{IvBur,Bur1,BurKer,BurMag,BurNst},
the complex time parameter $\t$ is cancelled in the Kerr
generating function $F$. Because of that, the world line $\oX
_{-}^\m$, having the same complex shift in the space-like
direction and the same 4-velocity, may also be used on the equal
reason and yields the same result.

The Kerr generating function $F$  may be represented in the form
\cite{IvBur,Bur1,BurKer,BurMag,BurNst}

\be F= (\lambda _{\dot\alpha} - \olam_{\dot\alpha})\check K\lambda
^{\dot\alpha} \ ,
 \label{genF} \ee

where \be \olam _{\dot\alpha}=\epsilon_{\dot\alpha
\dot\beta}\oX_{+}^{\m} \bar\sigma_\m^{\dot\beta\alpha} \psi_\alpha
\ . \label{olam}\ee are the values of twistor parameters at the
complex world line of the Kerr-Newman source $\oX_{+}^{\m}$.
Operator $\check K = u^\m\d _\m$ is related to momentum of the
complex particle and expressed, in accordance with  (\ref{umu}),
via vectors $k_L$  and $k_R$. Setting the equivalence of these
vectors for the Kerr-Newman and Dirac particles, one matches their
spin and momentum.

Without loss of generality one can set $\Re e \oX _{+}^\m =0.$
 Taking into account the relations (\ref{s10}) and (\ref{3m})
one obtains for the commuting spinors

\be \olam ^{\dot\alpha}= -2ia (\bar\chi \psi) \chi ^{\dot\alpha} ,
\ee

and

\be K\lambda ^{\dot\alpha}=\Pi^\m \bar\sigma_\m \psi  = -\frac m
{(\bar\phi\chi)} [(\phi\psi) \bar \phi ^{\dot\alpha} + (\bar\chi
\psi)  \chi ^{\dot\alpha}] . \ee

As a result,  function $F$ acquires  the  form (up to the
nonessential factor $m$)

\bea F(\psi, x^\m)= x^\m[(\phi\psi)(\bar\phi
\bar\sigma_\m\psi)+(\bar\chi\psi)(\chi \bar\sigma_\m
\psi)]/(\bar\phi\chi)   \nonumber   \\
- 2ia (\phi\psi) (\bar\chi \psi) \label{Fdirac} . \eea

One sees that it is the function which possesses all the
properties which are necessary for the Kerr generating generating
function. It is holomorphic in $\psi$ in accord with the
conditions of the Kerr theorem, and it is homogenous in $\psi$,
which allows one to transform it to the standard Kerr-Schild form
by the replacement $Y=\psi_2/\psi_1.$ Finally, it is quadratic in
$\psi$, which corresponds to the studied before case yielding the
Kerr congruence up to the shifts and Lorentz transformations
\cite{IvBur,KerWil,BurMag,BurNst}.

The Dirac spinor solutions $\phi, \chi$  depend via (\ref{Pi})
from the vector potential $A_\m$ which  is an external
electromagnetic field. Therefore, the Dirac wave function
$\Psi=(\phi, \chi)$ plays in this model the role of an order
parameter which controls dynamic  of the Dirac -- Kerr-Newman
particle, spin-polarization, momentum  and deformation of the Kerr
congruence caused by external electromagnetic field.

On the other hand, {\it the behavior of the Dirac -- Kerr-Newman
spinning particle turns out to be fully determined by the Dirac
wave function and will be indistinguishable from the behavior of
Dirac electron,} at least in the weak and slowly changed on the
Compton lengths electromagnetic field. It allows also to use the
standard stochastic interpretation of the wave function $\Psi (x)$
corresponding to one-particle quantum theory.

\section{Renormalization by gravity and regularization of
self-energy}

The mass renormalization is the most complicate and the most
vulnerable procedure in QED.
 Gravitational field is ignored in QED, relying on the argument
that its local action is negligible. However, gravity has a strong
non-local action which automatically provides the self-energy
renormalization for island sources.

Indeed, mass of an isolated source is determined by only
asymptotic gravitational field, and therefore, it depends only on
the mass parameter $m$ which survives in the asymptotic expansion
for the metric. On the other hand, the total mass can be
calculated as a volume integral which takes into account densities
of the electromagnetic energy $\rho_{em},$ material (mechanical
mass) sources $\rho_{m}$ and energy of gravitational field
$\rho_{g}$. The last term is not taken into account in QED, but
namely this term provides perfect renormalization. For a
spherically symmetric system, the expression may be reduced to an
integral over radial distance $r$\fn{It looks like the expressions
in a flat space-time. However, in the Kerr-Schild background, the
exact Tolman relations \cite{Tol}  taking into account
gravitational field and rotation give just the same result
\cite{BEHM}.}

\be m = 4\pi \int_0^\infty \rho_{em} dr + 4\pi\int_0^\infty
\rho_{m} dr + 4\pi\int_0^\infty \rho_{g} dr \label{mtot} . \ee

Some of these terms may  be divergent, but the total result will
not be changed, since divergences always will be compensated by
contribution from gravitational term.  Therefore, in the island
systems {\it gravity performs perfectly the mass-energy
renormalization for arbitrary distributions  of the charges and
matter.}

It shows that, due to the strong non-local action, gravity turns
out to be essential for elementary particles, on the distances
which are very far from the Planck scale.

The Kerr-Schild form of metric allows one to consider a broad
class of regularized solutions which remove the Kerr singular
ring, covering it by a matter source. There is a long-term story
of the attempts to find some interior regular solution for the
Kerr or Kerr-Newman solutions \cite{Isr,Lop,Man,BurBag,BEHM}.
\fn{Extra references may be found in \cite{Man,BurBag,BEHM}.}
Usually, the regularized solutions have to retain the Kerr-Schild
form of metric \be g_\mn=\eta_\mn +2H k_\m k_n \ee and the form of
Kerr principal null congruence $k_\m(x),$ as well as its property
to be geodesic and shear-free. The space part $\vec n$ of the Kerr
congruence $k_\m=(1,\vec n)$ has the form of a spinning hedgehog.
In the case $a=0,$ it takes the form of the spherically symmetric
hedgehog which is usually considered as an ansatz for the
solitonic models of elementary particles and quarks. It suggests
that this model may also have relation to the other elementary
particles. Indeed, the Kerr-Schild class of metric has a
remarkable property, allowing us to consider a broad class of the
charged and uncharged, the spinning and spinless solutions from an
unified point of view.

Our treatment will be based on the approach given in
\cite{BurBag,BEHM}, where the {\it smooth} regularized sources
were obtained for the rotating and non-rotating solutions of the
Kerr-Schild class.\fn{It was also published in the proceedings of
conferences, in particular \cite{RenGra}.} These smooth and
regular solutions are based on the G\"ursey and G\"urses ansatz
\cite{GG}, in which the Kerr-Schild scalar function $H$ of the
general form \be H=f(r)/(r^2 + a^2 \cos^2 \theta) \label{hf}. \ee
For the Kerr-Newman solution function $f(r)$ has the form \be
f(r)\equiv f_{KN}= mr -e^2/2 \label{hKN}. \ee

Regularized solutions have tree regions:

i) the Kerr-Newman exterior, $r>r_0 $, where $f(r)=f_{KN},$

ii) interior $r<r_0-\delta $, where $f(r) =f_{int}$ and function
$f_{int}=\alpha r^n ,$ and $n\ge 4$ to suppress the singularity at
$r=0,$ and provide the smoothness of the metric up to the second
derivatives.

iii) a narrow intermediate region $r\in [r_0-\delta, r_0]$ which
allows one to get a smooth solution interpolating between regions
i) and ii).

It is advisable to consider first the non-rotating cases, since
the rotation can later be taken into account by an easy trick. In
this case, taking $n=4$ and the parameter $\alpha=8\pi \Lambda/6
,$ one obtains for the source (interior) a space-time of  constant
curvature $R=-24 \alpha$ which is generated by a source with
energy density

\be \rho = \frac 1 {4\pi} (f'r -f)/\Sigma^2 , \label{rhof} \ee

and tangential and radial pressures

\be p_{rad}=-\rho, \quad p_{tan}=\rho - \frac 1 {8\pi}f''/\Sigma ,
\label{p}\ee

where $\Sigma=r^2.$ It yields for the interior the stress-energy
tensor $ T_{\mn} = \frac {3\alpha} {4\pi} diag (1,-1,-1,-1), $ or

\be\rho=-p_{rad}=-p_{tan}=\frac {3\alpha} {4\pi}, \label{rho} \ee

which generates a de Sitter interior for $\alpha >0,$ anti de
Sitter interior for $\alpha <0$. If $\alpha =0, $ we have a flat
interior which corresponds to some previous classical models of
electron, in particular, to the Dirac model of a charged sphere
and to the L\'opez model in the form of a
 rotating elliptic shell \cite{Lop}.

The resulting sources may be considered as the bags filled by a
special matter with positive ($\alpha>0$) or negative ($\alpha <
0$) energy density.\fn{It resembles the discussed at present
structure of dark energy and dark matter in Universe. The case
$\alpha >0$ is reminiscent of the old Markov suggestions to
consider particle as a semi-closed Universe \cite{FMM}.}

The transfer from the external electro-vacuum solution to the
internal region (source) may be considered as a phase transition
from `true' to `false' vacuum in a supersymmetric $U(1) \times
\tilde U(1) $ Higgs model \cite{BurBag,RenGra}.

Assuming that transition region iii) is very thin, one can
consider the following graphical representation which turns out to
be very useful.
\begin{figure}[ht]
\centerline{\epsfig{figure=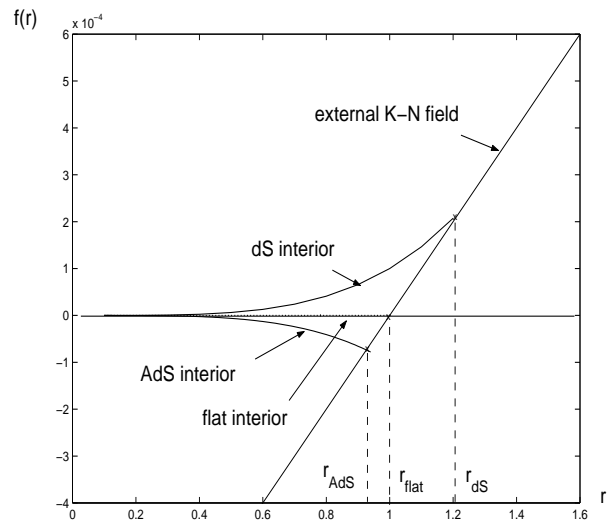,height=7cm,width=8cm}}
\caption{Regularization of the Kerr-Newman spinning particle by
matching the external field with  dS, flat or AdS interior.}
\end{figure}

The point of phase transition $r_0$ is determined by the equation
$f_{int}(r_0)=f_{KN}(r_0)$ which yields $\alpha r_0^4 = mr_0
-e^2/2 .$ From (\ref{rho}), we have $\rho=\frac {3\alpha} {4\pi}$
and obtain the equation

\be m= \frac {e^2} {2r_0} + \frac 4 3 \pi r_0^3 \rho . \ee

 In the first term on the right side, one can easily recognize
 the electromagnetic mass of a charged sphere with radius $r_0$,
 $M_{em}(r_0)=\frac {e^2} {2r_0}$, while the second
 term is the mass of this sphere filled by a material with a
 homogenous
 density $\rho$, $M_m =\frac 4 3 \pi r_0^3 \rho .$
 Thus, the point of intersection $r_0$ acquires
a deep physical sense, providing an energy balance by the mass
formation. In particular, for the classical Dirac model of a
charged sphere with radius $r_0=r_e=\frac {e^2} {2m},$ the balance
equation yields the flat internal space with $\rho=0.$ If $r_0>
r_e$, a material mass of positive energy $M_m>0$ gives a
contribution to total mass $m$. If $r_0 < r_e ,$ this contribution
has to be negative $M_m <0 , $ which is accompanied by the
formation of an AdS internal space.

All the above treatment retains valid for the rotating cases, and
for the passage to a rotating case, one has to set

\be \Sigma=r^2 +a^2 \cos^2 \theta , \ee and consider $r$ and
$\theta$ as the oblate spheroidal coordinates.

The Kerr-Newman spinning particle with a spin $J=\frac n 2 \hbar,$
 acquires the form of a relativistically rotating disk.
The corresponding stress-energy tensor (\ref{rho}) describes in
this case the matter of source in a co-rotating with this disk
coordinate system. Disk has the form of a highly oblate ellipsoid
with thickness $r_0$ and radius $a=\frac n 2 \hbar/m$ which is of
order the Compton length. Interior of the disk represents a
`false' vacuum having superconducting properties, so the charges
are concentrated on the surface of this disk, at $r=r_0$. Inside
the disk the local gravitational field is negligible.

\section{Non-stationarity and regularization of the
zero-point field.}

Classical models of spinning particle encounter an unavoidable
contradiction with quantum theory.

Metric of the Kerr-Newman solution is only a stationary
approximation for the metric which may be related to spinning
particle. In the old Kerr's microgeon model \cite{Bur0}, the Kerr
singular ring acquires electromagnetic wave excitations which were
interpreted as excitations of a circular string
\cite{Bur1,IvBur,BurOri}. Such excitations break axial symmetry of
the Kerr-Newman solution and stationarity. As a result, only an
average metric takes the Kerr-Newman form.
 Integration of the Kerr-Schild equations in the general non-stationary
 case \cite{DKS} leads to the appearance of an extra electromagnetic
field depending on some function $\gamma(x)$ which describes
electromagnetic radiation along the Kerr congruence $k_\m$
\cite{BurAxi}. In \cite{DKS}, authors set the restriction
$\gamma=0$ to obtain the final, stationary form of the Kerr-Newman
solution. The exact Einstein-Maxwell solutions with $\gamma\ne0$
have not been obtained so far and represent the old and very hard
problem.\fn{The wave electromagnetic solutions on the Kerr-Newman
background which are asymptotically exact in the low-frequency
limit where obtained recently in \cite{BEHM1}.} There are known
the Vaidya ``shining star'' solutions \cite{KraSte,BurNst}, in
which $\gamma \ne 0 ,$ but the Maxwell equation have been switched
out there, since the non-coherent radiation is only considered.
These solutions show that radiation leads also to the
non-stationarity via the loss of mass.

Note, that in the one-particle quantum theory oscillations are
stationary and absence of radiation caused by oscillations is
postulated. Contrary to quantum physics, in classical theory
non-stationarity entails radiation, and here lies a rather sharp
boundary between the classical and quantum theories.

It should be mentioned that radiation is present in QED too, since
the structure of free particles in QED is related to radiative
corrections caused by  vacuum field of virtual photons:  zero
point field and vacuum polarization.

By application of the quantum field theory in the curved spaces
\cite{deWit,BirDev}, quantum effects are concentrated in the
 stress-energy tensor which is divergent due to the vacuum zero point
 fields. By a
transfer to the classical Einstein-Maxwell theory, the quantum
vacuum fields have to be subtracted, which means that the
classical stress-energy tensor has to be regularized with respect
to the vacuum fluctuations \cite{deWit}

\be T^{(reg)}_{\mn} = T_{\mn} - <0|T_{\mn}|0>, \ee

under the condition

\be T^{(reg) \ \mn} ,_\m = 0 . \label{cons}\ee

Since the non-stationarity of the Kerr-Schild solutions is related
to the field $\gamma$, it was conjectured
\cite{Bur1,IvBur,BurTwi,BurNst} that this field has to be
subtracted by some regularization. Indeed, the field $\gamma$ is
an electromagnetic radiation propagating along the Kerr congruence
$k_\m$, and it has to involve a loss of mass. However, {\it
twofoldedness} of the Kerr geometry leads to a very specific
effect: {\it the outgoing radiation on the ``positive'' out-sheet
of the metric is compensated by an ingoing radiation on the
``negative'' in-sheet , and therefore, the solution has to be
stationary, as if the loss of mass is absent.}  It shows, that the
field $\gamma$ has to be identified with the vacuum zero-point
field and it has to be subtracted from the stress-energy tensor by
means of a procedure of regularization which has to satisfy the
condition (\ref{cons}). Such regularization may be performed,
\cite{Bur1,BurOri}, and leads to some modified Kerr-Schild
equations (see Appendix C). Although the equations are essentially
simplified, there is still a remnant of the field $\gamma$ in the
Maxwell equations, which reflects a relation between the real
electromagnetic excitations and vacuum fields. By such a
regularization, electromagnetic excitations may be interpreted as
a resonance of the zero-point fluctuations on the
(superconducting) source of the Kerr spinning particle
\cite{Bur1,BurOri,BurPra}.

 Although, the exact nontrivial solutions of the regularized system
were not obtained so far, there were obtained corresponding exact
solutions of the Maxwell equations which showed that any
excitation of the Kerr geometry leads to the appearance of some
extra ``axial'' singular line (string) which is semi-infinite and
modulated by de Broglie periodicity
\cite{BurAxi,BurOri,BurTwi,BurPra}. \fn{It was argued in
\cite{BurTwi,BurPra}, that the ``axial'' strings may acquire the
quark indices and are responsible for the scattering at high
energies.} There are exact Kerr-Schild solutions of this type
containing the ring-like singular string and semi-infinite
``axial" string. They corresponds to the end points of the complex
Kerr-Newman string and appear as the real images of the complex
point-like ``quarks'' which were mentioned in the end of section
II. The obtained recently multiparticle Kerr-Schild solutions
\cite{Multiks} supported this point of view, showing that
interaction occurs via pp-strings of this type, which we shall
discuss in the next section.

\section{Dirac-Kerr electron and multiparticle Kerr-Schild solutions.}

Some evidences that the Dirac -- Kerr-Newman model is related to a
multi-particle representation may be extracted from the obtained
recently multiparticle Kerr-Schild solutions. It was shown in
\cite{Multiks} that taking the generating function of the Kerr
theorem $F$ in the form of the product of partial functions for
i-th particle

\be F=\prod _i F_i(Y|q_i) \label{multiF}, \ee

where $q_i$ is the set of parameters of motion and orientation of
particle i, one can obtain the multi-particle Kerr-Schild
solutions of the Einstein-Maxwell system in the assumption that
particles are stationarily moving along some different
trajectories.

 The main equation of the Kerr theorem for twistorial structure
(\ref{F}) is satisfied by any partial solution $F_i(Y)=0$. It
means that
 the twistorial multi-particle space-time splits on the sheets
 corresponding to different roots of the equation $F(Y)=0,$
similar to the sheets of a Riemann surface.

Twistorial structures on the different sheets turn out to be
independent and twistorial structure of $i$-th particle ``does not
feel" the structure of particle $j,$ forming a sort of its
internal space. This property is direct generalization of the
corresponding property of the Kerr-Newman geometry - the
twosheeted structure of space-time, which splits  up the `in' and
`out' fields.

Since function $F(Y)$ for one Kerr-Newman particle is quadratic in
$Y$ \cite{IvBur1,KerWil,BurKer,BurNst}, the equation
$F_i(Y|q_i)=0$ has two roots $Y_i^+$ and $Y_i^-$ corresponding to
the positive (`out') and negative (`in') sheets. In terms of these
roots one can express $F_i$ in the form \cite{Multiks}

\be F_i(Y)=A_i(x)(Y-Y_i^+)(Y-Y_i^-).\label{Fi}\ee

One sees that metric of a multi-particle solution will depend on
the solution $Y_i (x)$ on the considered sheet of i-th particle.
Indeed,
 substituting the $(+)$ or $(-)$ roots $Y_i^\pm (x)$ in the
relation (\ref{Y}), one determines the Kerr congruence
$k^{(i)}_{\mu}(x)$ and corresponding function $h_i$ of the
Kerr-Schild ansatz
 (\ref{KS}) on the i-th sheet

 \be H_i = \frac m{2} (\frac
 1{\mu_i\tilde r_i} + \frac 1{\mu_i^*\tilde r_i^*}) + \frac {(e/\mu_i)^2}{2
 |\tilde r_i|^2} . \label{hKSi} \ee

Electromagnetic field generated by electric charge is given by the
vector potential \be A_\m^{(i)} =\Re e (\frac e {\mu_i\tilde r
_i}) k^{(i)}_\m \label{Aisol} .\ee

 The complex radial distance $\tilde r_i$ and function $\mu _i(Y)$ are
 also determined from the extended version of the Kerr theorem
\cite{Multiks},
 \be \tilde r_i= - d_Y F_i , \ee
 \be \mu_i (Y_i)= \prod _{j\ne i} A_j
 (x)(Y_i - Y_j^+) (Y_i - Y_j^-) \label{mui}. \ee

 Contrary to independence of twistorial structures for different particles,
there is an interaction between them, since the function $\mu
_i(Y)$ acquires a pole $\mu _i \sim A(x) (Y_i^+ - Y_j^-)$ on the
twistor line which is common for the particles $i$ and $j$. The
metric and electromagnetic field will be
 singular along the common twistor lines.
 For example, a light-like interaction occurs
along the line which connects the $out $ - sheet of particle $i$
to the $ in $ - sheet of particle $j$ , see fig.3. In the Witten
twistor-string model \cite{Wit,Nair} such a pole is considered as
propagator for a free fermion.

\begin{figure}[ht]
\centerline{\epsfig{figure=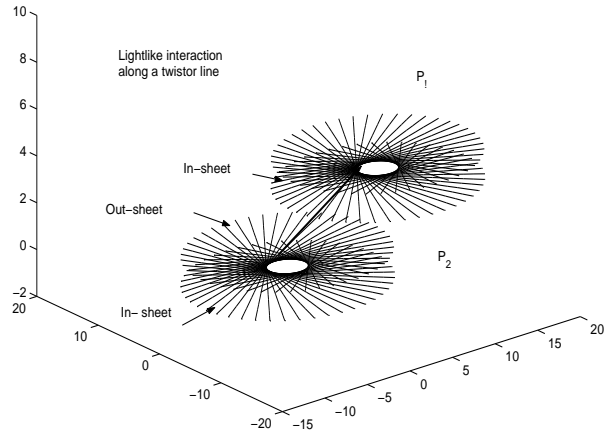,height=7cm,width=8cm}}
\caption{The light-like interaction
 via a common twistor line connecting  out-sheet of one particle
 to in-sheet of another.}
\end{figure}
This singular line is extended to infinity, since it will also be
common for the $out$ - sheets of the both particles. The field
structure of this line is similar to singular pp-wave solutions
\cite{KraSte,BurAxi}. Analysis of some simple cases shows that
each particle has a pair of semi-infinite singular lines which is
caused by interaction with some external particle. \fn{The
light-like strings contain only the
 one-way light-like modes. On the other hand, the modes of opposite
directions link the positive sheet of the second particle to
negative sheet of the first one, which is a stringy analog of the
photon exchanges.} As it was discussed in
\cite{BurTwi,BurPra,BurAxi} such a pair of strings turns out to be
a carrier of de Broglie periodicity.

\begin{figure}[ht]
\centerline{\epsfig{figure=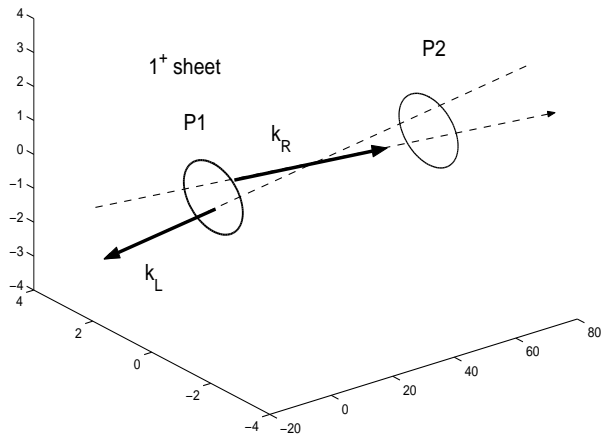,height=7cm,width=8cm}}
\caption{Two outgoing semi-infinite singular lines of the particle
P1, caused by its interaction with external particle P2.}
\end{figure}

In the limit of infinitely many external particles,
 the selected Kerr's particle will be connected by singular twistor
lines with many other external particle, and the singular twistor
lines will have even dense distribution among the twistor lines,
covering the principal null congruence of the selected particle.

\begin{figure}[ht]
\centerline{\epsfig{figure=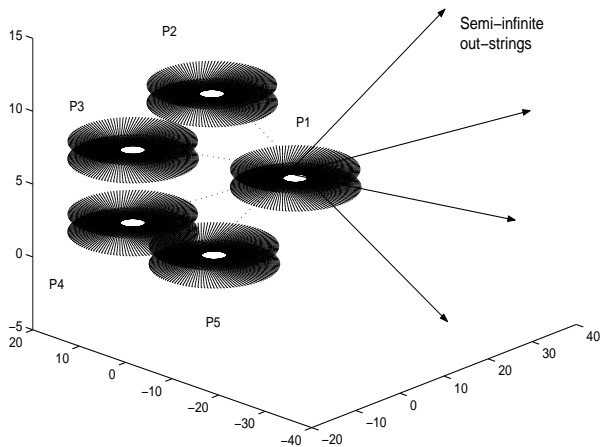,height=7cm,width=8cm}}
\caption{Formation of the outgoing semi-infinite singular lines by
interaction via the common `out-sheet'-`out-sheet' twistor lines.}
\end{figure}

The multiparticle Kerr-Schild solutions show us that the usual
Kerr-Newman solution is the solution for an isolated particle,
while there is a series of the exact corresponding solutions, in
which the selected Kerr particle is surrounded by other particles,
perhaps on the very far distances. This new type of the
Kerr-Newman solution differs from the initiate one in one respect
only - some of the twistor lines of the Kerr principal null
congruence turn out to be {\it singular}. These singular lines are
the light-like Schild strings which are described by singular
pp-waves \cite{BurTwi,BurAxi}. It shows, that the Dirac-Kerr model
is multiparticle indeed, because of the stringy inter-particle
interactions,  and also that the singular twistor lines are
analogs of the virtual photons which provide vacuum polarization
and radiative corrections by the mass and charge renormalization
in QED \cite{BLP}. Therefore, we arrive at the conclusion that the
Dirac -- Kerr-Newman
 electron takes an intermediate position between the one-particle
Dirac model and  the multiparticle  QED.

Returning now to the discussed in sec.VII regularization of the
stress-energy tensor for the wave solutions,
 one can see that the multiparticle Kerr-Schild solutions confirm the
 above conjecture that the wave field $\gamma$ propagating along the Kerr
congruence have to be identified with the  field of virtual
photons and have to be removed by regularization.
 On the other hand, it allows one to suppose that {\it
 the singular pp-strings of the multiparticle Kerr-Schild
 solutions represent elements of the vacuum structure,
 suggesting that vacuum has apparently a twistor-string
 texture.}

\section{Discussion: The wave function and space-time structure
of electron}

{\bf The `point-like' and `extended' electron.}

One of the main consequences of the considered here Dirac --
Kerr-Newman model is that electron acquires a definite space-time
structure of the Kerr-Newman geometry. In particular, it acquires
a twistorial structure having a caustic on the Kerr-Newman ring of
the Compton size, and the wave excitations of this ring-like
string are related with the wave properties of electron
\cite{BurTwi,BurPra}.

On the other hand, the Dirac equation controls the motion of the
Kerr spinning particle, and consequently, the behavior of this
model does not differ from the corresponding behavior of the Dirac
electron, in particular, in the external electromagnetic field. It
is known that the external field in the one-particle Dirac model
has to be restricted by the conditions to be small enough and
weakly changing on the Compton distances \cite{BLP,BjoDre,AkhBer}.
These are exactly the conditions, which allow us to consider the
Dirac -- Kerr-Newman spinning particle as a united system. Note
that similar restrictions exist also for the multi-particle
treatments in QED, as the conditions for decomposition on the
creation and annihilation operators. Therefore, the model may be
interpreted in terms of the one-particle Dirac theory which
contains, however, a hidden space-time structure, and in this
respect it is also close to the local QED.

 In accordance with QED, the naked electron is point-like, while the
 dressed by virtual photons electron is deformed and
 turns out to be smeared over the Compton region
\cite{Schw,BLP}. Therefore, calculation of this effect in QED is
based on the Coulomb solution which is used as a first
approximation in coordinate representation \cite{BLP}. It is clear
that these calculations do not take into account the spin of
electron which may be described in the first approximation  by the
Kerr-Newman field. Although the Kerr-Newman gravitational field is
very weak and  may apparently be neglected, the corresponding
electromagnetic field cannot be considered as small. The extremely
high spin of electron polarizes the space-time, and the aligned to
the Kerr congruence electromagnetic field turns out to be
polarized along the Kerr-Newman twistorial structure, acquiring
singularity at the ring of Compton radius. This effect of
polarization has also to be acting for the field of virtual
photons. Therefore, the use of the Kerr-Newman electromagnetic
field instead of the Coulomb field would be more correct in this
case. It is clear that the resulting field of virtual photons
shall again be concentrated near the Kerr-Newman singular ring of
the Compton size forming its stringy excitation.

In is known that the relativistic quantum theory, in spite of the
great success in the experimental confirmations, has brought some
problems of theoretical consistency \cite{BLP,BjoDre,AkhBer}. In
particular, the Dirac wave solutions cannot be considered
consequently as the wave functions, since the relativistic quantum
theory cannot be considered consistently as one-particle theory
\cite{BLP,BjoDre,AkhBer}. There appears the problem of
localization of electron and stochastic interpretation of the
Dirac wave function. The problems exhibited manifestly in
coordinate representation, leading experts to the conclusion (see
\cite{BLP}) that ``... the wave functions $\psi(q)$ as the
carriers of unobservable information cannot be used in the
consequent relativistic theory ...''. There appears even the
extreme points of view that the modern relativistic quantum theory
has to refuse to deal with the wave function  at all and also with
the treatment of processes in space-time, and that the Feinman
graphs and S-matrix may represent itself the final form of the
quantum theory \cite{BLP,BjoDre}.

The considered here combined Dirac -- Kerr-Newman model of
electron realizes  point of view that the spinning particle has
some distinct extended space-time structure which is represented
in a Lorentz-covariant form and does not contradict to the main
conclusions of the one-particle Dirac theory and to the structure
of QED. Besides, the combined Dirac -- Kerr-Newman model turns out
to be naturally related with gravitational field.

{\bf Wave function as an order parameter}

The presented here combined Dirac - Kerr-Newman model may be
considered in the frame of the usual one-particle quantum theory
with the interpretation of the Dirac wave function as a density of
probability for the position of electron. There appears however an
extra local information on the space-time polarization for each
point $x\in M^4 ,$ which is determined by the Dirac wave function
$\Psi(x).$ This information may also be interpreted in the  spirit
of the local quantum field theory, in which the role of wave
function may be more essential.

 In sec. III we obtained that there is a natural
null tetrad $ n^a $ related to the Dirac wave function. This
tetrad is a direct generalization of the null tetrad  $ \{ d \z, \
d \Z, \ du, \ dv \}$ which is determined by the Cartesian null
coordinates and plays important role in the Kerr-Schild formalism
and twistor theory. The spin-direction of the Kerr's particle in a
standard representation is determined by the real direction
$du+dv$ which has for the Dirac electron the analog $(k_{L\m} -
k_{R\m})dx^\m $. Similar, the complex forms $d\z, \ d\Z$ are
analogs of the Dirac null vectors $m_\m dx^\m$ and $\bar m_\m
dx^\m$. These complex vectors of the Dirac tetrad have a very
important peculiarity. They are oscillating with the double
Compton frequency and are carrying de Broglie periodicity for the
moving electron.

There is especial interest to use corresponding generalization of
the Kerr-Schild formalism to the case of oscillating tetrad.
Formally, the choice of the tetrad is only a gauge transformation
which will not result to any change in the nature of the
corresponding solutions of the Einstein-Maxwell system of
equation. However, two factors are important:

i - it may simplify the equation, opening a way for obtaining some
new solutions (for example oscillating ones) which were not
obtained so far in the Kerr-Schild formalism, and

ii - there may be a topologically non-trivial situation, in which
the new tetrad will not be equivalent to initiate one. An
important example of this situation is the twist of
spin-structure\fn{Do not mix with twist of the Kerr congruence.}
which mixes the internal and orbital degrees of freedom and has
found important applications in the solitonic models, topological
sigma-models and twistor-string theory \cite{Wit,Nair}.

In this case, the Dirac wave function acquires again the role of
an order parameter (Goldstone fermion) controlling the broken
symmetry which is related  to fixation of the tetrad. Therefore,
in the Dirac -- Kerr-Newman model,  the Dirac wave function may
acquire the role of an `order parameter' which controls the
twistorial polarization of space-time, fixing
 the broken symmetry of the surrounding vacuum.

It should also be noted, that the twistorial structure of the
Dirac -- Kerr-Newman model allows one to transform the Dirac wave
function from coordinate representation to the projective twistor
space. The twistor-space representation turns out to be more
natural for the Kerr-Newman geometry than the both coordinate and
momentum representations. Taking into account that S-matrix and
scattering amplitude can naturally be represented in twistor space
too, as it has been shown for the more simple case of gauge
scattering amplitudes in twistor-string model \cite{Wit}, there
appears apparently some new approach to QED which  could resolve
the existing problems with the wave function in coordinate
representation.

\section{Conclusion}

The considered Dirac -- Kerr-Newman model of electron has a few
attractive properties:

- electron acquires an extended space-time structure which
does-not contradict to conclusions of the Dirac model and QED,

- the Kerr-Newman twistorial structure is controlled by the Dirac
equation, so all the consequences of this model have to match to
the reach experimental data obtained from the Dirac equation, and
since the Feinman rules are also based on the Dirac equation, one
can expect that this model will not conflict with the results of
QED too,

 - the model shows that gravity performs
renormalization and regularization of the particle-like models in
a very elegant manner,

- the model has a non-trivial geometrical structure which displays
a relationship to quantum theory. It opens a way to a geometrical
view on the formal quantum prescriptions.

A peculiarity of this model is that the Dirac equation takes an
especial role with respect to the other fields - it is  considered
on the auxiliary Minkowski space-time, while the other fields are
considered on the Kerr-Schild background. One could argue that
this difference is not essential, since the local gravitational
field of
 electron is extremely small and will not affect on the solutions of the
 Dirac equation.  However, it is not correct, indeed.
 The Kerr solution has the
 nontrivial twofold topology which is retained even in the limit of
 the infinitely small mass. Because of that
 the aligned to the Kerr congruence electromagnetic  and
 spinor  solutions, which could lead to the self-consistent solutions,
 turn out to be essentially different from the plane wave solutions on
 Minkowski space-time, which is illustrated by the
  appearance of the extra axial singular filament. So, the plane
wave solutions are deformed with concentration
 of the fields near singular lines.

In the considered Dirac -- Kerr-Newman model the Dirac wave
function plays the especial role of a `master equation' which
controls the polarization and set a synchronization of tetrad. The
perspective models of another sort, in which the Dirac equation
`feels' the Kerr geometry, i.e. has a back reaction from the Kerr
geometry, may be based on the stringy structures of the Kerr
geometry. It has to use some initially massless Dirac solutions on
the Kerr space-time, forming a foliation over the complex Kerr
string. The non-zero mass term of the Dirac equation appears in
this model by averaging over the string length, similar to the
appearance of the mass in the massless dual string models. We
expect to consider such a model elsewhere.
\section*{Acknowledgments}
This work was supported by the RFBR Grant 07-08-00234 and by the
ISEP Research Grant by Jack Sarfatti. Author thanks for
stimulating discussion G. Alekseev, A. Efremov, L. Pitaevskii, V.
Ritus and O. Teryaev, and also the participants of the seminar on
Quantum Field Theory held by M. Vasiliev at the Lebedev Physical
Institute. Author is also thankful to J. Bagger for the useful
correspondence on derivation of spinor relations.

\section*{Appendix A: Spinor notations}

\be \gamma^\m  = \left( \begin{array}{cc}
0 & \sigma ^\m \\
\bar\sigma ^\m & 0
\end{array} \right) \ ,
\ee where \be \bar\sigma ^{\m \dot\alpha \alpha} = \epsilon ^{\dot
\alpha \dot \beta} \epsilon ^{ \alpha  \beta} \sigma ^\m _{\beta
\dot\beta}, \label{bsigma} \ee and
\begin{equation}
\begin{array}{cc}
\sigma _0  = \bar\sigma _0 = \left( \begin{array}{cc}
1 & 0 \\
0 & 1
\end{array} \right) \ ,
\quad \sigma _1  = -\bar\sigma _1 = \left( \begin{array}{cc}
0 & 1 \\
1 & 0
\end{array} \right) \ , \\  \nonumber \\
\sigma _2  = - \bar\sigma _2 = \left( \begin{array}{cc}
0 & -i \\
i & 0
\end{array} \right) \ ,
\quad \sigma _3  = - \bar\sigma _3 = \left( \begin{array}{cc}
1 & 0 \\
0 & -1
\end{array} \right) \ .
\label{sigma}
\end{array}
\end{equation}
$\epsilon^{12}=1, \quad \epsilon^{21}=\epsilon_{12}=-1 .$

Also,

\be \gamma ^0 = \left( \begin{array}{cc}
0 & -1 \\
-1 & 0
\end{array} \right),
 \ee

\be \gamma_5=\gamma ^5=\gamma^0 \gamma^1 \gamma^2 \gamma^3 =\left(
\begin{array}{cc}
-i & 0 \\
0 & i
\end{array} \right) .
 \ee

\section*{Appendix B: Some relations for commuting spinors}

\be\epsilon_{\alpha\beta} \epsilon ^{\beta\gamma} = {\delta
_\alpha}^\gamma ,\label{s2}\ee

\be\epsilon^{12}=\epsilon_{21}=1, \quad
\epsilon^{21}=\epsilon_{12}=-1 .\label{eps}\ee

\be\psi_\alpha \chi ^\alpha = \epsilon_{\alpha\beta}\psi^\beta
\epsilon ^{\alpha\gamma} \chi _\gamma = -
\psi^\beta\epsilon_{\beta\alpha} \epsilon ^{\alpha\gamma} \chi
_\gamma =- \psi^\beta \chi_\beta =- \psi\chi \ . \label{s4}\ee

It yields

 \be\psi\chi =-\psi_\alpha\chi^\alpha
 = -\chi^\alpha \psi_\alpha=
 -\chi\psi \Rightarrow \psi\psi=0 \ ,
 \ \ \ \ \ \ \label{s6}\ee

Complex conjugation changes the order of spinors without change of
sign, which yields

\be(\chi\psi)^+ = (\chi^\alpha \psi_\alpha)^{*T}
=(\bar\chi^{\dot\alpha}\bar\psi_{\dot\alpha})^{T} =
(\bar\psi_{\dot\alpha}\bar\chi^{\dot\alpha}) , \ee

and  due to (\ref{s6})

\be(\chi\psi)^+ = (\chi^\alpha \psi_\alpha)^+ =\bar\psi
_{\dot\alpha} \bar\chi^{\dot\alpha} =\bar\psi \bar\chi =-\bar\chi
\bar\psi. \ \ \label{s8}\ee

Next, using the relation \be \sigma_{\alpha\dot\alpha}=
\epsilon_{\alpha\beta} \epsilon _{\dot\alpha\dot\beta}
\bar\sigma^{\dot\beta\beta} \ee and taking into account that
$\epsilon \psi =-\psi \epsilon$, one obtains

\be\chi ^\alpha \sigma _{\alpha\dot\alpha}\bar\psi ^{\dot\alpha} =
\chi _\beta \bar\psi _{\dot\beta}\bar\sigma^{\dot\beta\beta} \ ,
\ee

which yields (assuming that $\chi^\alpha \to \chi _\alpha$and
$\bar\psi^{\dot\alpha}\to \bar\psi_{\dot\alpha}$)

\be\chi\sigma\bar\psi =  \bar\psi \bar\sigma \chi  \ . \ \
\label{s10}\ee

Basing on the relation \be \sigma_{\alpha\dot\alpha}^\m
\bar\sigma^{\dot\beta\beta}_\m  = -2 \delta_\alpha^{ \ \beta}
\delta_{\dot\alpha}^{ \  \dot \beta} , \ee and taking into account
the order of the co- and contra-variant spinors, one  obtains

\be (\bar\chi \sigma^\m \phi)(\bar\sigma_\m \psi)^{\dot\alpha} =
 -2(\bar\chi\psi) \phi^{\dot\alpha} ,
\label{3m} \ee

and \be(\psi\phi)\bar\chi_{\dot\beta}= -\frac 12
(\phi\sigma^\m\bar\chi) (\psi\sigma_\m)_{\dot\beta} \ . \ \
\label{s102}\ee

We have also
 \be (\chi\sigma\bar\psi)^+ = \psi \sigma
\bar\chi \ , \ \ \ \ \label{s12C}\ee

and

\be (\chi\sigma^\m \bar\sigma^\n \psi)^+ = \bar\psi \bar\sigma^\n
\sigma ^\m \bar\chi \ . \ \ \label{s13C}\ee

\section*{Appendix C: Regularized Einstein-Maxwell equations}

The field equations for the Einstein-Maxwell system after
preliminary integration in  \cite{DKS} are the following.
Electromagnetic field is given by tetrad components of the
self-dual tensor $\cF _{ab}$, \be \cF _{12} =AZ^2 \label{1}, \ee
\be \cF _{31} =\gamma Z - (AZ),_1  \ , \label{2}\ee and the
equations relating the functions $A(x)$ and $\gamma (x)$ are \be
A,_2 - 2 Z^{-1} \cZ Y,_3 A  = 0 , \label{3}\ee \be \cD A+  \cZ
^{-1} \gamma ,_2 - Z^{-1} Y,_3 \gamma =0 . \label{4}\ee Here
$Z=P/\tilde r $ and $ Y,_3 = -ZP_Y/P $ (notations of \cite{DKS}).

Gravitational equations take the form \be M,_2 - 3 Z^{-1} \cZ Y,_3
M  = A\bar\gamma \cZ ,  \label{5}\ee \be \cD M  = \frac 12
\gamma\bar\gamma  , \label{6}\ee where \be \cD=\d _3 - Z^{-1} Y,_3
\d_1 - \cZ ^{-1} \Y ,_3 \d_2   \ . \label{cD} \ee Solutions of
this system were given in \cite{DKS} only for the case for
$\gamma=0$. Regularization of the stress-energy tensor
 \be T_{reg}^\mn = \ :T^\mn: \ \equiv T^\mn
-<0|T^\mn|0> \ee under the condition $\nabla _\m T_{reg}^\mn =0$
retains the term $\gamma$ in the Maxwell equations (\ref{1}),
(\ref{2}),  (\ref{3}), (\ref{4}) and in the gravitational equation
(\ref{5}), while the equation (\ref{6}) takes the simple form \be
\cD M  = 0  , \label{6a}\ee which provides stationarity.

\end{document}